\RequirePackage{ifpdf}
\ifpdf 
\documentclass[pdftex]{sigma}
\else
\documentclass{sigma}
\fi

\begin{document}
\allowdisplaybreaks

\renewcommand{\PaperNumber}{015}

\FirstPageHeading

\ShortArticleName{Auger Spectra and Dif\/ferent Ionic Charges}

\ArticleName{Auger Spectra and Dif\/ferent Ionic Charges Following
$\boldsymbol{3s}$, $\boldsymbol{3p}$ and $\boldsymbol{3d}$
Sub-Shells Photoionization of Kr Atoms}

\Author{Yehia A. LOTFY~$^\dag$ and Adel M. El-SHEMI~$^\ddag$}
\AuthorNameForHeading{Y.A. Lotfy and A.M. El-Shemi}

\Address{$^\dag$~Physics Department,  Faculty of Science, El Minia University,\\
$\phantom{^\dag}$~P.O. Box  61111, El Minia, Egypt}
\EmailD{\href{mailto:yahialotfy59@yahoo.com}{yahialotfy59@yahoo.com}}

\Address{$^\ddag$~Applied Sciences Department, College of
Technological Studies,\\
$\phantom{^\ddag}$~P.O. Box 42325, Shuwaikh, 70654 Kuwait}
\EmailD{\href{mailto:admohamed@yahoo.com}{admohamed@yahoo.com}}

\ArticleDates{Received August 21, 2005, in f\/inal form January 15,
2006; Published online January 31, 2006}

\Abstract{The decay of inner-shell vacancy in an atom through
radiative and non-radiative transitions leads to f\/inal charged
ions. The de-excitation decay of $3s$, $3p$ and $3d$ vacancies in
Kr atoms are calculated using Monte--Carlo simulation method. The
vacancy cascade pathway resulted from the de-excitation decay of
deep core hole in $3s$ subshell in Kr atoms is discussed. The
generation of spectator vacancies during the vacancy cascade
development gives rise to Auger satellite spectra. The last
transitions of the de-excitation decay of $3s$, $3p$ and $3d$
holes lead to specif\/ic charged ions. Dirac--Fock--Slater wave
functions are adapted to calculate radiative and non-radiative
transition probabilities. The intensity of ${\rm Kr}^{4+}$ ions
are high for $3s$ hole state, whereas ${\rm Kr}^{3+}$ and ${\rm
Kr}^{2+}$ ions have highest intensities for $3p$ and $3d$ hole
states, respectively. The present results of ion charge state
distributions agree well with the experimental data.}


\Classification{81V45}

\section{Introduction}

The relaxation of inner-shell ionized atom via successive Auger,
Coster--Kronig and radiative transitions leads to production of
highly charged ions. In the course of de-excitation decay pathway,
multiple vacancies are generated after each Auger transition. The
distribution of generated vacancies does not dependent on the
initial ionization process. The Auger cascades are accompanied
with emission of photon spectra or electron spectra for each
pathway branch. The spectra are conditioned by the transition
rates and transition energies of multi-vacancy states. The
generation of vacancies in the course of de-excitation cascade is
accompanied by characte\-ristic energy shifts in the electronic
levels. The inf\/luence of the additional vacancies during the
cascades may close some low-energy Auger channels (forbidden
energies). Understanding the inf\/luence of the spectator vacancies
on Auger transitions, gives more detailed information about the
vacancy cascade development. The overlapping spectra emitted from
parallel branches of the de-excitation cascades lead to dif\/ferent
low-energy highly charged ions. These low-energy ionic charges are
important in the f\/ield of astrophysical plasma
\cite{McDowell&Ferendici}. Study of ion charge state distributions
following inner-shell ionization of atoms provides information to
estimate the characteristic relaxation time constants and thermal
equilibrium of ion gas that stored in trap
\cite{Church&Kravis1,Church&Kravis2,Kravis&Church}.  The ion
charge state distributions following inner-shell ionized rare gas
atoms are measured using both energy ionization from $x$-ray tube
\cite{Carlson&Krause,Krause&Carlson}, and synchrotron radiation
\cite{Ueda&Shigemasa,Hayaishi&Morioka,Mukoyama&Tonuma,Saito&Suzuki,Tawara&Hayaishi}.
Tamenori et al.~\cite{Tamenori&Okada} measured the branching ratio
of multiply charged ions formed through photoionization of Kr
$3d$, $3p$ and $3s$ subshells using a coincidence technique. The
calculations of ion charge state distributions resulted after
core-hole creation of atom are carried out by many researchers
\cite{Omare&Hahn1,Omare&Hahn2,Kochur&Dudenko1,Kochur&Dudenko2,Opendak,Mukoyama,Mirakhmedove&Parilis,
El-Shemi&Ghoneim,Abdullah&El-Shemi&Ghoneim,El-Shemi&Lotfy}.

The Auger transition rates and transition energies in
multi-ionized neon atom are measured using heavy-ion and electron
bombardment \cite{Matthews&Johnson} and calculated using
Hartree--Fock--Slater wave functions \cite{Bhalla&Folland}.
Larkins \cite{Larkins} investigated the inf\/luence of multi-vacancy
states on the electron binding energies and
    transition rates in K, L, and M subshells in the calculation of ion charge
    state distributions. Auger and $x$-ray spectra formed during vacancy cascades are
    calculated using Monte--Carlo method \cite{Mirakhmedov,Mirakhmedov&Parilis}. Cooper
    et al.~\cite{Cooper&Southworth&MacDonald} measured and
    calculated the L$_{23}$MM Auger spectra of argon excited at energies between those
    of K- and L-thresholds and af\/fected by ${\rm L}_{1}{\rm L}_{23}$M intermediate transition.
     The L$_{23}$MM Auger spectra of argon emitted after photoexcitation are measured
      using broad-band synchrotron radiation of energies largely lying above the K-shell
       ionization potential~\cite{VonBusch&Doppelfeld&Gunther}.

    In the present work, the dif\/ferent ionic charges followed $3s$, $3p$ and
$3d$ subshells photoionization of Kr are calculated using Monte
Carlo simulation technique. The calculation is considered an
extension to the previous work carried out by Abdullah et
al.~\cite{Abdullah&El-Shemi&Ghoneim}, in which ion charge state
distributions yield after $1s$, $2s$ and $2p$ vacancy creation in
Kr atoms are obtained. The de-excitation decay branches of
parallel Coster--Kronig and Auger transitions followed $3s$, $3p$
and $3d$ hole production are discussed. The Auger transition rates
and electron shake of\/f are obtained using Dirac--Fock--Slater wave
functions. The present results are compared with available
theoretical \cite{Kochur&Dudenko2} and experimental values
\cite{Tamenori&Okada}.

\section{Method of calculations}

The highly excited atoms resulted after electron emission from
inner-shell will rapidly relax back to a lower energy state by
radiative or non-radiative process. In case of radiative
transition, the vacancy transfers to a higher shell under emission
of characteristic $x$-rays:
\[
A+h\nu\longrightarrow (A^{1+})^\ast+e_{p}\longrightarrow
(A^{1+})^\ast\longrightarrow A^{1+}+h\nu',
\]
where the core hole in the target atom A is produced by
photoionization $(h\nu)$, $(A^{1+})^{\star}$ is the atom in highly
excited state and $e_{p}$ is the primary emitted electron. If the
vacancy is f\/illed via Auger transition, two vacancies in the
higher shells of atom will be created and the transition is given
by:
\[
A+h\nu\longrightarrow (A^{1+})^\ast+e_{p}\longrightarrow
(A^{1+})^\ast\longrightarrow A^{2+}+e_{\rm Auger},
\]
where $A^{2+}$  is doubly ionized atom.

In an Auger transition, one electron falls from a higher level to
f\/ill the initial vacancy and another electron is ejected into
continuum. The kinetic energy of emitted Auger electron can be
estimated from the binding energies of the various levels involved
in the transition:
\[
E_{Kin}=E_{i}-(E_{j}+E_{k}),
\]
where $E_{i}$ is the binding energy of initial state, $E_{j}$ is
the binding energy of electron f\/illing the hole and $E_{k}$ is the
binding energy of the electron leaving the atom.

 The simulation of
the cascade branches is based on the selection of all possible
radiative and non-radiative branching ratios that may f\/ill the
inner-shell vacancies in atoms. The radiative branching ratio is
def\/ined as the probability that the vacancy in an atom is f\/illed
through $x$-ray transitions (photon emission), while the
non-radiative branching ratio is def\/ined as the vacancy f\/illed
through Auger and Coster--Kronig processes. The radiative and
non-radiative branching ratio are given by:

 For f\/luorescence yield:
\[
\omega(f\longrightarrow i)=\Gamma^{R}_{if}/\Gamma
\]
and for Auger yield:
\[
\omega(f\longrightarrow i)=\Gamma^{A}_{if}/\Gamma,
\]
where $i$ is the initial conf\/iguration decaying into the f\/inal
conf\/iguration $f$. $\Gamma$  is the sum of partial radiative
widths $\Gamma^{R}_{if}$ and non-radiative width $\Gamma^{A}_{if}$
which is given by:
\[
\Gamma=\sum_{i,f}\Gamma^{R}_{if}+\sum_{i,f}\Gamma^{A}_{if}.
\]
 The calculations of radiative transition rates are performed for
  singly ionized atoms using mul\-ticonf\/iguration-Dirac--Fock (MCDF) wave
   fun\-c\-tions~\cite{Grant&Mckenzie&Norrington}. The non-radiative transition rates and electron shake
   of\/f processes are computed using Dirac--Fock--Slater (DFS) wave func\-tions~\cite{Lorenz&Hartmann}.

The Monte--Carlo simulation technique is described in detail by
Abdullah et al.~\cite{Abdullah&El-Shemi&Ghoneim}, in which the
electron shake of\/f process is considered. In radiative transitions
case, the inner-shell vacancy will transfer to higher shell under
emission of characteristic $x$-rays without changing the number of
vacancies thus a new position of vacancy is created. If the
vacancy is f\/illed via Auger process, two vacancies in the higher
shells of atom will be generated. The de-excitation decays are
repeated until all vacancies reach the outermost shell and no
transitions are possible. The successive Auger cascades lead to
the production of specif\/ied low-energy charged ions. The Auger
cascades correspond to transitions with emission of photon spectra
or electron spectra for each pathway branches. In the course of
cascade development, Auger spectra arise from spectator vacancies
left by preceding Auger transitions and by electron shake of\/f
process accompanying the Auger transitions.

\section{Results and discussions}
Fig.~1 shows a schematical diagram for all main de-excitation
decay pathway resulted after~$3s$ vacancy creation in Kr atom. The
solid arrows indicate the selected Auger channels follo\-wing
inner-shell ionization, while dotted ones show the Coster--Kronig
channels. The solid line indicates the radiative transition; while
dotted lines indicate the electron shake of\/f processes occuring
due to the change of atomic potential after primary ionization or
after Auger and Coster--Kronig transitions. Electron shake of\/f
process resulted from the initial ionization or~de-excitation
decay produces additional vacancies in higher shells leading to an
increase in the number of vacancies. Auger and Coster--Kronig
transitions are indicated outside the brackets, while the
spectator vacancy conf\/igurations are indicated inside the
brackets. Each branch of the de-excitation leads to an ion of a
specif\/ic charge as shown in the diagram. Considering of electron
shake of\/f process in the calculation gives a f\/inal charge state of
ions which agree well with the experimental data.

\begin{figure}[t]
\centerline{\begin{minipage}{7.0cm}
\centerline{\includegraphics[width=7cm]{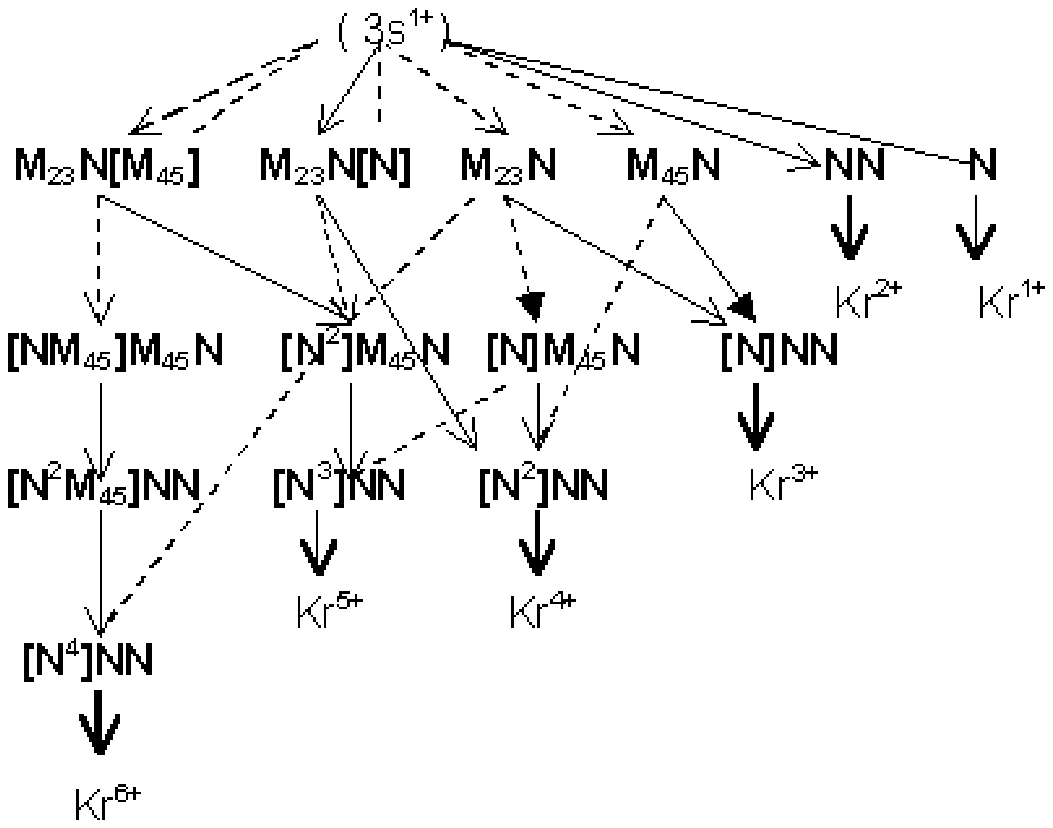}}
\vspace{5mm} \caption{Decay branches pathway after $3s$ vacancy
production in Kr atom.}
\end{minipage}
\qquad
\begin{minipage}{7.0cm}
\centerline{\includegraphics[width=6.4cm]{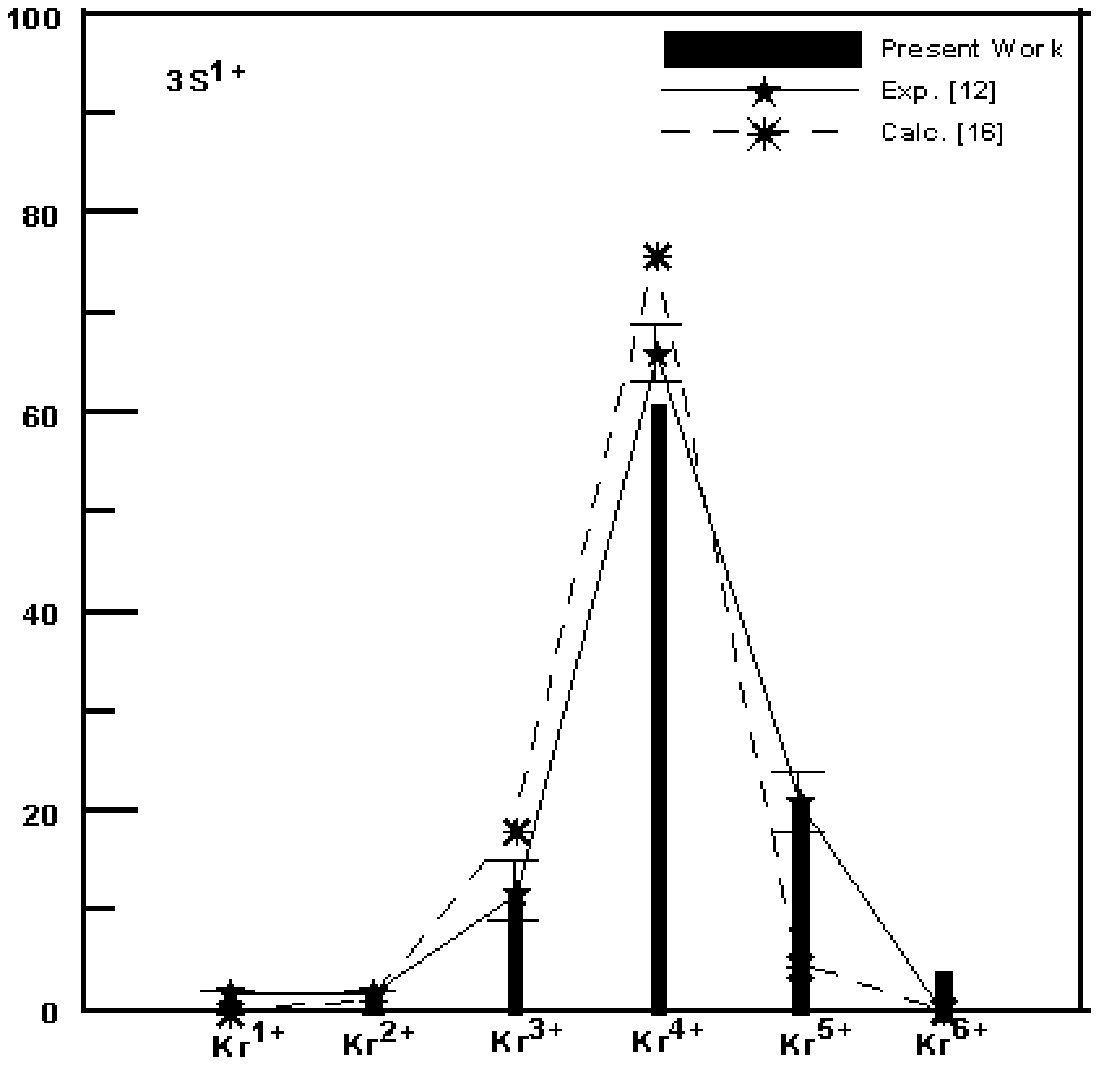}} \vspace{-3mm}
\caption{Dif\/ferent ionic charges formed following $3s$ shell
ionization in Kr atom.}
\end{minipage}}
\vspace{-4mm}
\end{figure}

  Fig.~2 shows the charge state of ions resulted after $3s$ vacancy
creation in Kr atoms. The present results are compared with other
theoretical values from Kochur et al. \cite{Kochur&Dudenko2} and
experimental data from Tamenori et al.~\cite{Tamenori&Okada}. The
relaxation of ionized atom occurs through radiative and/or
non-radiative transitions. At $3s$ hole state in Kr atoms, the
probability of radiative transitions is lower than the probability
of non-radiative transitions. The radiative branching ratios will
occur in 1\,\% of the cases as given by the f\/luorescence yield. In
the remaining 99\,\% of the cases, the $3s$ vacancy is f\/illed by
non-radiative processes (Auger and Coster--Kronig transitions).
The ionization of the $N$ valence electron through radiative
transitions leads to the formation of ${\rm Kr}^{1+}$ ions. The
intensity of singly charged ions followed $3s$ hole creation is
less than~1\,\%. The $3s$ hole is f\/illed by one of three possible
${\rm M}_{1}{\rm M}_{23}{\rm N}$, ${\rm M}_{1}{\rm M}_{45}N$
Coster--Kronig transitions and ${\rm M}_{1}{\rm NN}$ Auger
transition. At $3s$ hole state, the probability of super
Coster--Kronig transitions ${\rm M}_{1}{\rm MM}$ is lower than the
probability of Auger and Coster--Kronig transitions. The decay of
$3s$ vacancy through Auger or Coster--Kronig channel will create
an additional vacancy in N subshells. The generation of~N vacancy
produced during vacancy cascade leads to characteristic energy
shifts in the energy levels. This additional N vacancy closes
low-energy super Coster--Kronig ${\rm M}_{1}{\rm MM}$ channels and
these channels become energy forbidden. The relative transition
 probability (branching ratio) for ${\rm M}_{1}{\rm M}_{23}{\rm N}$, ${\rm M}_{1}{\rm M}_{45}{\rm N}$ and
 ${\rm M}_{1}{\rm NN}$ decays
are 72\,\%, 26\,\% and 1.5\,\%, respectively. The intensity of
doubly charged ions ${\rm Kr}^{2+}$ at $3s$ hole state is about
2\,\%. The yield of ${\rm Kr}^{3+}$ ions with intensity of 12\,\%
arises from Coster--Kronig channels and subsequent Auger
transitions following $3s$ hole creation. The formation of ${\rm
Kr}^{4+}$ is caused by initial ${\rm M}_{1}{\rm M}_{23}{\rm N}$
Coster--Kronig transition and subsequent Coster--Kronig transition
from ${\rm M}_{23}{\rm M}_{45}{\rm N}$ shell and ${\rm M}_{45}{\rm
NN}$ Auger transition. During the vacancy cascade de-excitation,
the intensity of ${\rm M}_{45}{\rm NN}$ Auger spectra with
additional $[{\rm N}^{2}]$ vacancies is high. The $3s$ ionization
signif\/icantly produces quadruply charged $({\rm Kr}^{4+})$ ions
via Coster--Kronig decays with subsequent Auger transitions. The
intensity of quadruply charged ions resulted from 3s vacancy is
61\,\%. The spectator vacancies [N] decrease the energy of ${\rm
M}_{45}{\rm NN}$ Auger transition by several electron volts. This
is because the spectator vacancies change the shielding of the M
and N levels. The strongest line spectrum arise from ${\rm
M}_{45}[{\rm N}^{2}]-{\rm NN}[{\rm N}^{2}]$ transitions and leads
to a stable ${\rm Kr}^{4+}$ ions. The ${\rm Kr}^{5+}$ and ${\rm
Kr}^{6+}$ ions are formed from the de-excitation via parallel
Auger and Coster--Kronig channels and electron shake of\/f process.
The intensities of the ${\rm Kr}^{5+}$ and ${\rm Kr}^{6+}$ ions
are 21\,\% and less than 4\,\%, respectively. It is found that the
creation of additional N vacancies during the de-excitation decay
of $3s$ hole closes the ${\rm M}_{23}{\rm M}_{45}{\rm M}_{45}$
super Coster--Kronig channel, i.e.\ the super Coster--Kronig
transition energy is forbidden. The intensity of ${\rm Kr}^{4+}$
and ${\rm Kr}^{5+}$ ions is high while a ${\rm Kr}^{1+}$, ${\rm
Kr}^{2+}$, ${\rm Kr}^{3+}$ and ${\rm Kr}^{6+}$ ions is low. The
most probable de-excitation pathway of a $3s$ shell vacancy is via
Coster--Kronig transitions, which will create ${\rm M}_{45}$
vacancies with additional N sub-shell vacancies. So each spectator
vacancy conf\/iguration will give rise to an ${\rm M}_{45}{\rm NN}$
satellite spectrum. The present results of ion charge state
distributions after $3s$ vacancy production agree well with the
experimental data \cite{Tamenori&Okada}.

\begin{figure}[t]
\centerline{\begin{minipage}{7cm}
\centerline{\includegraphics[width=6.4cm]{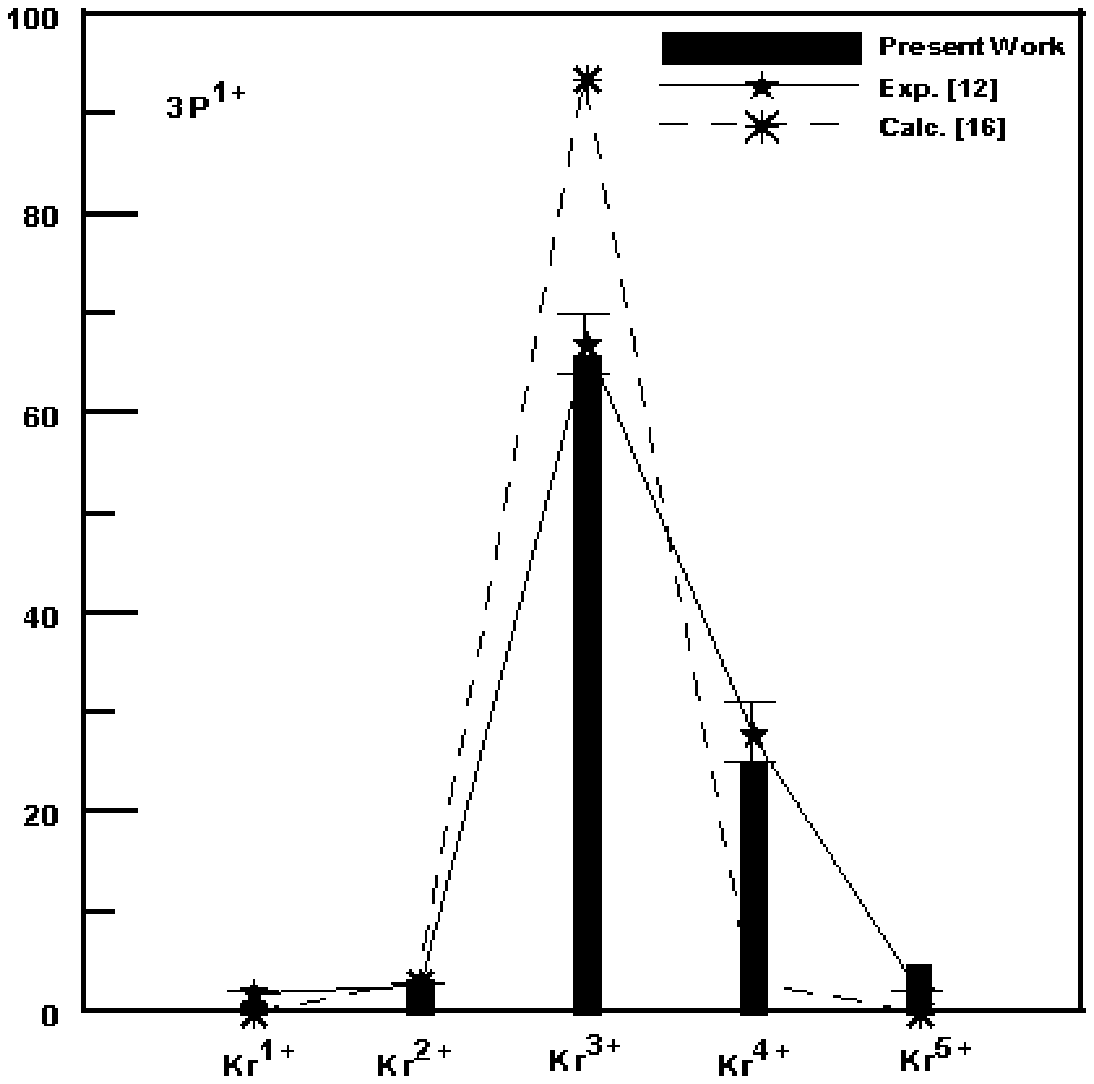}} \vspace{-3mm}
\caption{Dif\/ferent ionic charged following $3p$ shell ionization
in Kr atom.}
\end{minipage}
\qquad
\begin{minipage}{7cm}
\centerline{\includegraphics[width=6.4cm]{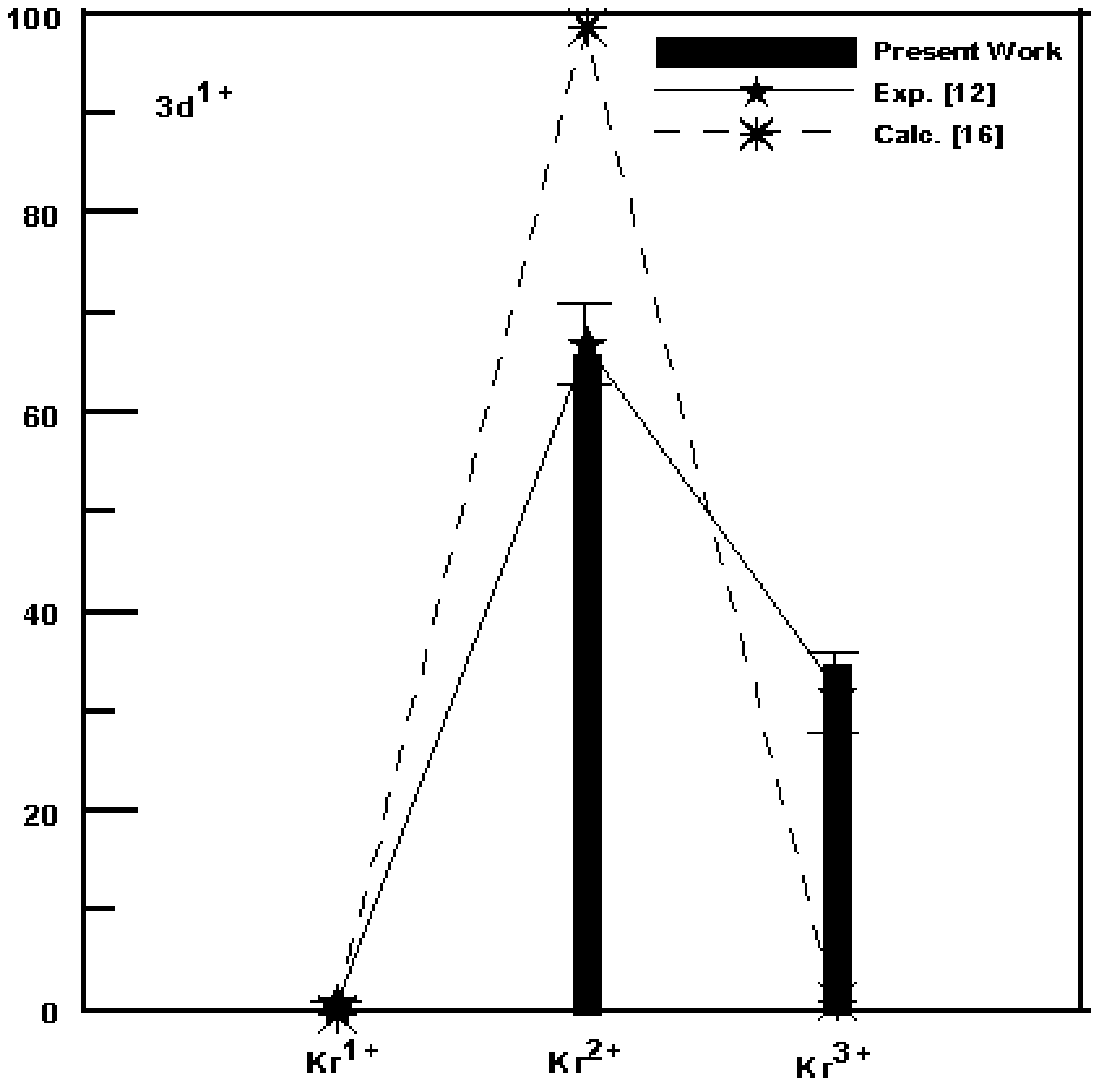}} \vspace{-4mm}
\caption{Dif\/ferent ionic charged following $3d$ shell ionization
in Kr atom.}
\end{minipage}}
\vspace{-3mm}
\end{figure}

The ion charge state distributions produced after $3p$ hole
creation in Kr are shown in Fig.~3. The de-excitation decay of
$3p$ hole gives rise to dif\/ferent ionic charges ${\rm Kr}^{q}$,
$q=1,\ldots,5$.  The $3p$ hole is f\/illed by one of the following
possible transitions: ${\rm M}_{23}{\rm M}_{45}{\rm M}_{45}$ super
Coster--Kronig, ${\rm M}_{23}{\rm M}_{45}{\rm N}$ Coster--Kronig
and ${\rm M}_{23}{\rm NN}$ Auger transitions. The relative
transition probability (bran\-ching ratio) for ${\rm M}_{23}{\rm
M}_{45}{\rm M}_{45}$, ${\rm M}_{23}{\rm M}_{45}{\rm N}$ and ${\rm
M}_{23}{\rm NN}$ decays are 61\,\%, 29\,\% and 2\,\%,
respec\-tively.   In the $3p$ hole state, the ionization of the $3p$
electrons increases the intensity of triply charged ions ${\rm
Kr}^{3+}$ and decreases the quadruply charged ions ${\rm Kr}^{4+}$
in comparison with the corresponding charged ions that produced
after $3s$ vacancy. The formation of ${\rm Kr}^{1+}$ is caused by
radiative transition with ionizing electron from N~subshells. The
intensity of ${\rm Kr}^{1+}$ ions after $3p$ shell ionization is
less than~1\,\%. The decay of $3p$ vacancy via normal ${\rm
M}_{23}{\rm NN}$ Auger transition leads to the formation of doubly
charged ions ${\rm Kr}^{2+}$ with an intensity of 3\,\%.
 The formation of ${\rm Kr}^{3+}$ ions occurs from the decay of $3p$ hole via initial
Coster--Kronig channel and subsequent Auger transition. The triply
charged ions ${\rm Kr}^{3+}$ followed after the $3p$ ionization
has highest intensity (66\,\%). The ${\rm Kr}^{4+}$ ions with
intensity of 25\,\% are formed via Coster--Kronig, Auger
transitions and electron shake-of\/f from N~shell. The probability
of electron shake of\/f from N~shell after $3p$ ionization is about
7\,\%. The intensity of ${\rm Kr}^{5+}$ ions yields from vacancy
cascade followed $3p$ hole production is 5\,\%.
 In general, the de-excitation decay of $3p$ hole of Kr leads to dominant production
  of ${\rm Kr}^{3+}$ and ${\rm Kr}^{4+}$ charged ions and lower production
of ${\rm Kr}^{2+}$ and ${\rm Kr}^{5+}$ ions, respectively.

    The percentage
fraction of the ion charge state distributions yields from
de-excitation decay of $3d$ hole in Kr are shown in Fig.~4. The
highest intensity ${\rm Kr}^{2+}$ ions formed from de-excitation
decays of $3d$ photoionization. The de-excitation decays of $3d$
vacancy yields ${\rm Kr}^{2+}$ with higher intensity and ${\rm
Kr}^{3+}$ ions with lower intensity in comparison with the
corresponding results from $3p$ vacancy. The intensities of doubly
and triply charged ions are 66\,\% and 35\,\%, respectively. It is
important to mention that the present results of ion charge state
distributions
 produced after $3s$, $3p$ and $3d$ vacancies production in Kr atoms agree well with the
 experimental data from Tamenori et al.~\cite{Tamenori&Okada}.

\section{Conclusions}
 The core hole creation followed by successive Auger,
Coaster--Kronig transitions and electron shake of\/f leads to
dif\/ferent ionic charge. Monte Carlo simulation method is performed
to calculate the highly charged ions after $3s$, $3p$ and $3d$
vacancies creation in Kr atom. The Auger and Coster--Kronig
branching ratio and electron shake of\/f process are calculated
using Dirac--Fock--Slater wavefunctions. The cascade of decay
pathway branches followed $3s$, $3p$ and $3d$ holes are discussed.
The ${\rm Kr}^{4+}$ ions have highest intensity in the $3s$ shell
ionization, while ${\rm Kr}^{3+}$ and ${\rm Kr}^{2+}$ ions have
the highest intensity in the $3p$ and $3d$ shell ionization,
respectively. The present results are compared with other
theoretical values \cite{Kochur&Dudenko2} and
experimental~\cite{Tamenori&Okada} and are found to agree well
with the experimental values.

\LastPageEnding

\end{document}